\documentclass{elsart5}

\usepackage{graphicx}

\usepackage{amssymb}

\begin{document}

\begin{frontmatter}

\title{Effect of Hund's rule coupling on SU(4) spin-orbital system}

\author{Hiroaki Onishi\corauthref{cor1}}
\ead{onishi.hiroaki@jaea.go.jp}
\corauth[cor1]{}
\author{Takashi Hotta}
\address{%
Advanced Science Research Center,
Japan Atomic Energy Agency,
Tokai, Ibaraki 319-1195, Japan}
\received{5 July 2006}
%\revised{}
%\accepted{}

%use optional labels to link authors explicitly to addresses:
%\author{}
%\address{}

\begin{abstract}
We investigate the ground-state property of
a one-dimensional two-orbital Hubbard model at quarter filling
by numerical techniques
such as the density-matrix renormalization group method
and the exact diagonalization.
When the Hund's rule coupling $J$ is zero,
the model is SU(4) symmetric.
In fact, both spin and orbital correlations have a peak at $q$=$\pi/2$,
indicating an SU(4) singlet state with a four-site periodicity.
On the other hand, with increasing $J$, it is found that
the peak position of the orbital correlation changes to $q$=$\pi$,
while that of the spin correlation remains at $q$=$\pi/2$.
We briefly discuss how the SU(4) symmetry is broken by $J$.
\end{abstract}

%%%%%%%%%use the \KEY command at the begin of keyword text%%%%%%%%%
\begin{keyword}
\PACS 75.45.+j
\sep 75.10.-b
\sep 75.40.Mg
\KEY SU(4) spin-orbital system
\sep Hund's rule coupling
\sep density-matrix renormalization group
\end{keyword}
%Please supply one or more relevant PACS-1996 classification codes 
%(http://publish.aps.org/PACS/96pacs.html) and about 5 keywords 
%of your own choice for indexing purposes. 
%You can see a list of already used keywords for JMMM at 
%http://authors.elsevier.com/JournalDetail.html?PubID=505704&Precis=KIND

%75.45.+j Macroscopic quantum phenomena in magnetic systems
%75.10.-b General theory and models of magnetic ordering
%75.40.Mg Numerical simulation studies

\end{frontmatter}

%%%%%%%%%%%%%%%%%%%%%%%%%%%%%%%%%%%%%%%%%%%%%%%%%%%%%%%%%%%%
% Introduction
%%%%%%%%%%%%%%%%%%%%%%%%%%%%%%%%%%%%%%%%%%%%%%%%%%%%%%%%%%%%

It has been widely recognized that
the interplay of spin and orbital degrees of freedom plays
a significant role in the emergence of exotic magnetism
in strongly correlated electron systems with orbital degeneracy.
In this context,
the highest symmetric SU(4) spin-orbital model has been one of the
subjects of much interests from a theoretical viewpoint.
In particular,
the one-dimensional model is Bethe ansatz solvable \cite{Sutherland1975},
and the combined quantum effects of spin and orbital
have been revealed by analytical and numerical investigations
\cite{Li1999,Yamashita1998,Frishmuth1999}.
Indeed, it is characteristic of an SU(4) singlet state that
correlation functions show critical behavior with a four-site periodicity
and the elementaly excitation is gapless.

The highest SU(4) symmetry originates in the situation,
where electrons hop only between the same types of orbitals
with equal amplitude and the Hund's rule coupling is ignored
in a two-orbital Hubbard model.
In a more realistic situation, however,
the Hund's rule coupling should break the SU(4) symmetry
down to SU(2)$_{\rm spin}$$\times$U(1)$_{\rm orbital}$
\cite{Yamashita1998,Lee2004,Xavier2006}.
To clarify the effect of such symmetry breaking,
Lee \textit{et~al.} have studied
an SU(4) Hubbard model perturbed by the Hund's rule coupling
by means of renormalization-group and bozonization methods,
and proposed that the spin gap opens
for an arbitrarily small Hund's rule coupling \cite{Lee2004}.

%%%%%%%%%%%%%%%%%%%%%%%%%%%%%%%%%%%%%%%%%%%%%%%%%%%%%%%%%%%%
% Model and Method
%%%%%%%%%%%%%%%%%%%%%%%%%%%%%%%%%%%%%%%%%%%%%%%%%%%%%%%%%%%%

In this paper,
we investigate spin and orbital correlations
in a one-dimensional two-orbital Hubbard model
with one electron per site.
The Hamiltonian is given by
\begin{eqnarray}
 H=
 &&
 t\sum_{i,\tau,\sigma}
 (d_{i\tau\sigma}^{\dag} d_{i+1\tau\sigma}+\mbox{h.c.})
 + U \sum_{i,\tau} \rho_{i\tau\uparrow} \rho_{i\tau\downarrow}
 \nonumber\\
 &&
 + U'\sum_{i,\sigma,\sigma'} \rho_{i\alpha\sigma} \rho_{i\beta\sigma'}
 + J \sum_{i,\sigma,\sigma'}
 d_{i\alpha\sigma}^{\dag} d_{i\beta\sigma'}^{\dag}
 d_{i\alpha\sigma'} d_{i\beta\sigma}
 \nonumber\\
 &&
 + J'\sum_{i,\tau \ne \tau'} 
 d_{i\tau\uparrow}^{\dag} d_{i\tau\downarrow}^{\dag}
 d_{i\tau'\downarrow} d_{i\tau'\uparrow},
\end{eqnarray}
where $d_{i\tau\sigma}$ is the annihilation operator for an electron
with spin $\sigma$ in orbital $\tau$(=$\alpha,\beta$) at site $i$,
$\rho_{i\tau\sigma}$=$d_{i\tau\sigma}^{\dag}d_{i\tau\sigma}$,
and $t$ is the hopping amplitude.
Hereafter, $t$ is taken as the energy unit.
$U$, $U'$, $J$, and $J'$ denote
intra-orbital Coulomb, inter-orbital Coulomb,
exchange (Hund's rule coupling),
and pair hopping interactions, respectively.
Note that $U$=$U'$+$J$+$J'$
due to the rotational invariance in the local orbital space,
and $J$=$J'$ due to the reality of the wavefunction \cite{Dagotto2001}.
When the Hund's rule coupling is zero,
i.e., $U$=$U'$ and $J$=$J'$=0,
the system possesses the SU(4) symmetry.

We analyze the model with 32 sites ($N$=32) in the open boundary condition
by the density-matrix renormalization group method \cite{White1992}.
The finite-system algorithm is employed with keeping 400 states per block
and the truncation error is estimated to be $5\times 10^{-6}$ at most.
We also use the Lanczos method
for the analysis of a four-site periodic chain.
In this paper,
we set $U'$$-$$J$=20 to consider the strong-coupling region,
and investigate the dependence on $J$.

%%%%%%%%%%%%%%%%%%%%%%%%%%%%%%%%%%%%%%%%%%%%%%%%%%%%%%%%%%%%
% Results
%%%%%%%%%%%%%%%%%%%%%%%%%%%%%%%%%%%%%%%%%%%%%%%%%%%%%%%%%%%%

In Fig.~1, we show our DMRG results for
the spin and orbital structure factors, defined by
\begin{eqnarray}
 S(q)&=&\sum_{j,k}e^{{\rm i}q(j-k)}\langle S_{j}^{z}S_{k}^{z} \rangle/N,
 \\
 T(q)&=&\sum_{j,k}e^{{\rm i}q(j-k)}\langle T_{j}^{z}T_{k}^{z} \rangle/N,
\end{eqnarray}
with
$S_i^z$=$\sum_{\tau}
(\rho_{i\tau\uparrow}$$-$$\rho_{i\tau\downarrow})/2$ and
$T_i^z$=$\sum_{\sigma}
(\rho_{i\alpha\sigma}$$-$$\rho_{i\beta\sigma})/2$.
At $J$=0,
$S(q)$ and $T(q)$ coincide with each other,
and we can observe a peak at $q$=$\pi/2$,
clearly indicating an SU(4) singlet state with a four-site periodicity,
which is consistent with the previous numerical work
for the SU(4) spin-orbital model \cite{Yamashita1998}.
On the other hand,
$S(q)$ and $T(q)$ exhibit distinct behavior with increasing $J$,
as shown in Figs.~1(b) and 1(c).
It is observed that the peak position of $T(q)$ changes
from $q$=$\pi/2$ to $q$=$\pi$,
since $T(\pi/2)$ is suppressed,
while $T(\pi)$ is enhanced due to the effect of $J$.
As for the spin correlation,
the peak position of $S(q)$ remains at $q$=$\pi/2$
even for finite values of $J$,
since the magnitude of $S(\pi/2)$ increases
in sharp contrast to the case of $T(q)$.

%%%%%%%%%%%%%%% Fig. 1 %%%%%%%%%%%%%%%%%%%%%%%%%%%
\begin{figure}[t]
\centering
\includegraphics[width=\linewidth]{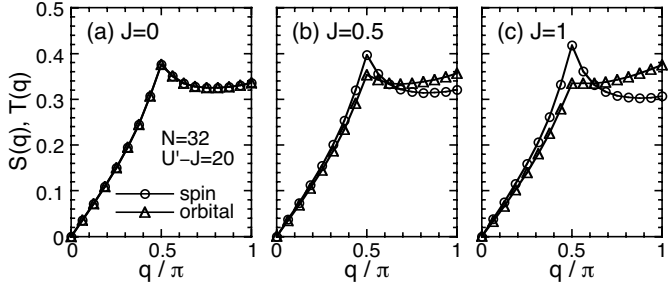}
\caption{%
Spin and orbital structure factors for
(a) $J$=0, (b) $J$=0.5, and (c) $J$=1.
}
\end{figure}
%%%%%%%%%%%%%%%%%%%%%%%%%%%%%%%%%%%%%%%%%%%%%%%%%%

To obtain intuitive understanding
for the changes in $S(q)$ and $T(q)$,
it is useful to consider a four-site system,
which is a minimal model to form the SU(4) singlet ground state
at $J$=0 \cite{Li1998}.
The SU(4) singlet state is expressed as
\begin{equation}
 \vert {\rm SU(4)} \rangle =
 (1/\sqrt{24}) \sum_{i \neq j \neq k \neq l}
 d_{i\alpha\uparrow}^{\dag}d_{j\alpha\downarrow}^{\dag}
 d_{k\beta\uparrow}^{\dag}d_{l\beta\downarrow}^{\dag}
 \vert 0 \rangle,
\end{equation}
where $\vert 0 \rangle$ is the vacuum state and
the summation is taken over all permutations for site indices.
Note that $\vert {\rm SU(4)} \rangle$ consists of
24 states with the same weight for each state.
For finite $J$, however,
the ground state is not represented by $\vert {\rm SU(4)} \rangle$ itself.
As shown in Figs.~2(a)-(c),
the 24 states are split into three classes
with eight states for each
according to the weight in the spin-singlet ground state \cite{Xavier2006}.
Note that each class is characterized by the peak positions of
$S(q)$ and $T(q)$,
denoted by $(q_{\rm spin},q_{\rm orbital})$:
$(\pi/2,\pi)$ for the class a,
$(\pi/2,\pi/2)$ for the class b, and
$(\pi,\pi/2)$ for the class c.

In Fig.~2(d), we show the $J$ dependence of the weight of
each class $m$ in the ground state,
\begin{equation}
 w_m=\sum_{i\in m}
 \vert \langle \phi_{i} \vert \psi_{\rm G} \rangle \vert^2,
\end{equation}
where $\psi_{\rm G}$ is the ground state
and $\phi_{i}$ denotes the basis.
At $J$=0, three classes contribute to the ground state
with equal weight.
With increasing $J$, the weight of the class a increases,
while those of the class b and c decrease and
the total weight of the three classes does not change.
In the class a, as shown in Fig.~2(a),
partly spin ferromagnetic (FM) alignment appears
due to the Hund's rule coupling,
and an antiferro-orbital (AFO) configuration is favored
to avoid the energy loss in the hopping process
due to the intra-orbital Coulomb interaction,
indicating the instability to a FM/AFO state
due to the Hund's rule coupling.
Thus, the correlations of $S(\pi/2)$ and $T(\pi)$ are enhanced,
leading to the change of the peak position of $T(q)$,
as shown in Fig.~1.
We note that with further increasing $J$,
the ground state is changed to a FM/AFO state.

We have tried to understand how the spin gap $\Delta_{\rm s}$
of the present model (1) is affected by $J$,
but it is a difficult task to estimate $\Delta_{\rm s}$ with
precision, because of charge, spin, and orbital degrees of freedom.
To clarify the behavior of $\Delta_{\rm s}$ in the thermodynamic limit,
it would be appropriate to investigate
an effective spin-orbital model in the strong-coupling limit,
which is an interesting future issue.

%%%%%%%%%%%%%%% Fig. 2 %%%%%%%%%%%%%%%%%%%%%%%%%%%
\begin{figure}[t]
\centering
\includegraphics[width=\linewidth]{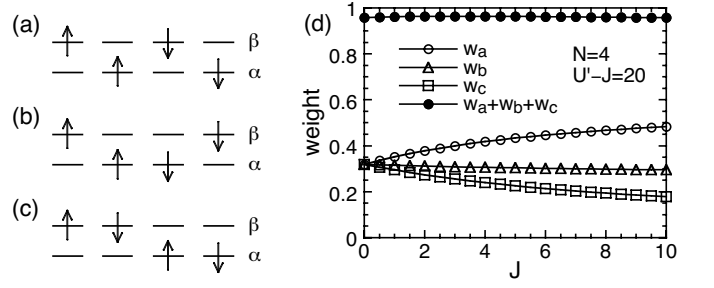}
\caption{%
(a)-(c) Three classes of electron configuration
in the SU(4) singlet state of the four-site system.
Each class has eight equivalent states
due to the translation and the change of orbitals.
(d) Weight in the ground state of each class in Figs.~2(a)-(c).
}
\end{figure}
%%%%%%%%%%%%%%%%%%%%%%%%%%%%%%%%%%%%%%%%%%%%%%%%%%

%%%%%%%%%%%%%%%%%%%%%%%%%%%%%%%%%%%%%%%%%%%%%%%%%%%%%%%%%%%%
% Summary
%%%%%%%%%%%%%%%%%%%%%%%%%%%%%%%%%%%%%%%%%%%%%%%%%%%%%%%%%%%%

In summary,
at $J$=0, $S(q)$ and $T(q)$ agree with each other
and have a peak at $q$=$\pi/2$ due to the SU(4) singlet ground state.
On the other hand, with increasing $J$,
we observe the transition of the peak position of $T(q)$ to $q$=$\pi$
in accordance with the change of relevant spin-orbital configuration.

%%%%%%%%%%%%%%%%%%%%%%%%%%%%%%%%%%%%%%%%%%%%%%%%%%%%%%%%%%%%
% Acknowledgment
%%%%%%%%%%%%%%%%%%%%%%%%%%%%%%%%%%%%%%%%%%%%%%%%%%%%%%%%%%%%

We thank K. Ueda for useful discussions.
T.H. is supported by the Japan Society for the Promotion of Science
and by the Ministry of Education, Culture, Sports, Science,
and Technology of Japan.

%%%%%%%%%%%%%%%%%%%%%%%%%%%%%%%%%%%%%%%%%%%%%%%%%%%%%%%%%%%%
% References
%%%%%%%%%%%%%%%%%%%%%%%%%%%%%%%%%%%%%%%%%%%%%%%%%%%%%%%%%%%%


\begin{thebibliography}{00}

%--- SU(4)
\bibitem{Sutherland1975}
B. Sutherland,
\textit{Phys. Rev. B} \textbf{12} (1975) 3795.

%--- SU(4)
\bibitem{Li1999}
Y. Q. Li, M. Ma, D. N. Shi, and F. C. Zhang,
\textit{Phys. Rev. B} \textbf{60} (1999) 12781.

%--- SU(4)
\bibitem{Yamashita1998}
Y. Yamashita, N. Shibata, and K. Ueda,
\textit{Phys. Rev. B} \textbf{58} (1998) 9114.

%--- SU(4)
\bibitem{Frishmuth1999}
B. Frischmuth, F. Mila, and M. Troyer,
\textit{Phys. Rev. Lett.} \textbf{82} (1999) 835.

%--- SU(4)+Hund
\bibitem{Lee2004}
H. C. Lee, P. Azaria, and E. Boulat,
\textit{Phys. Rev. B} \textbf{69} (2004) 155109.

%--- t2g Hubbard chain
\bibitem{Xavier2006}
J. C. Xavier, H. Onishi, T. Hotta, and E. Dagotto,
\textit{Phys. Rev. B} \textbf{73} (2006) 014405.

%--- manganite review
\bibitem{Dagotto2001}
E. Dagotto, T. Hotta, and A. Moreo,
\textit{Phys. Rep.} \textbf{344} (2001) 1.

%--- DMRG method
\bibitem{White1992}
S. R. White,
\textit{Phys. Rev. Lett.} \textbf{93} (1992) 2863.

%--- SU(4)
\bibitem{Li1998}
Y. Q. Li, M. Ma, D. N. Shi, and F. C. Zhang,
\textit{Phys. Rev. Lett.} \textbf{81} (1998) 3527.

\end{thebibliography}
\end{document}